\documentstyle[prc,aps,preprint]{revtex}
\def\3{\u{g}}
\def\5{{\i}}
\def\*{\"{u}}
\def\7{\"{o}}
\begin{document}
\draft
\date{\today}
\title{Associative Photoproduction of Roper Resonance and $\omega$-Meson}
\author{H. Babacan\,$^1$ and T. Babacan\,$^2$}
\address{$^1$\,Physics Department, Middle East Technical University,
06531 Ankara, Turkey\\
$^2$\,Physics Department, Celal Bayar University, Ya\3c\5lar
Kamp\*s\*, Manisa, Turkey } \maketitle
\begin{abstract}

Associative photoproduction of $\omega$-meson and $N^{*}(1440)$ on
nucleons, $\gamma+N\rightarrow\omega+N^{*}(1440)$, in the near
threshold region is investigated in a framework employing
effective Lagrangians. Besides $\pi$-exchange in t-channel, baryon
exchanges, i.e. N- and $N^{*}$-exchanges, in the s- and u-channels
are also taken into account in calculations of differential
cross-section and beam asymmetry. Important inputs of this model
are the vector and tensor coupling constants of $\omega
NN^{*}(1440)$-vertex, which are assumed to be equal to the values
of these couplings for $\omega NN$ vertex. Using our previous
estimation of $\omega NN$ coupling constants obtained from a fit
to available experimental data on photoproduction of $\omega$
meson in the near threshold region, we produce the necessary
numerical predictions for different observables in
$\gamma+N\rightarrow\omega+N^{*}(1440)$. Numerical results shows
that at low $|$t$|$ dominant contribution comes from t-channel
$\pi$-exchange while effects of nucleon and $N^{*}(1440)$ pole
terms can be seen at large $|$t$|$. Our predictions for the
differential cross section and beam asymmetry for the processes
$\gamma+N\rightarrow\omega+N^{*}(1440)$, where N is proton and
neutron at $E_{\gamma}=2.5$ GeV are presented with zero width
approximation and also with the inclusion of width effects of
$N^{*}(1440)$.
\end{abstract}
\section{Introduction}

Baryon resonances, in particular the Roper resonance
$N^{*}_{1/2,1/2}(1440)$, have a special interest at the moment
from the theoretical and experimental point of view. $N^{*}(1440)$
is the first excited state of nucleon with a broad full width of
$350\pm 100$ MeV which is twice as compared to those of the
neighbooring resonances $N^{*}(1520)$ and $N^{*}(1535)$ \cite{R1}.
Although the Roper resonance was discovered first during the phase
shift analysis of $\pi$-N scattering \cite{R2}, it has not
observed directly yet. Some quark \cite{R3,R4,R5,R6,R7} and bag
models \cite{R8,R9} try to explain its nature, but it is still not
well known. Photoproduction of vector mesons in particular
channel, where the target nucleon is excited to Roper resonance,
$\gamma+N\rightarrow\ V+N^{*}(1440)$, might provide supplementary
knowledge about this resonance and its couplings to meson-nucleon
channels.

In order to extract information on Roper resonance from assocative
production of vector meson and $N^{*}(1440)$ it is essential to
understand the production mechanisms. Since Roper resonance has
similar quantum numbers (spin 1/2, isospin 1/2 and positive
parity) with nucleon the corresponding dynamics of the associative
photoproduction of vector meson and $N^{*}(1440)$,
$\gamma+N\rightarrow\ V+N^{*}(1440)$, can be studied analogously
to that of "elastic" vector meson photoproduction,
$\gamma+N\rightarrow\ V+N$. Theoretical studies on photoproduction
of neutral vector meson
\cite{R10,R11,R12,R13,R14,R15,R16,R17,R18,R19} involve the
different combination of the following mechanisms: (i)
pseudoscalar ($\pi, \eta$) and scalar ($\sigma$)- meson exchanges
in t-channel; (ii) One-nucleon exchange in (s+u)-channel; (iii)
the Pomeron exchange in t-channel.

In Ref. \cite{R20}, associative neutral vector meson ($\rho$ and
$\omega$) and $N^{*}(1440)$ production near threshold in $\gamma
p$ interaction has been analyzed within an approach based on the
tree level diagrams of t- channel $\pi$- and $\sigma$- exchanges
and effective Lagrangians. For such exchanges although it is
possible to obtain some constraints on $\pi NN^{*}(1440)$ and
$\sigma NN^{*}(1440)$ couplings, measurement of above reactions
with linearly polarized photon will be more decisive. Considering
these both mechanisms alone will also result in trivial
polarization phenomena which can be predicted without knowledge of
exact values of coupling constants and phenomenological form
factors. For example, the beam asymmetry $\Sigma$ induced by the
linear polarization of the photon beam, and all possible T-odd
polarization observables as such, for example, target asymmetry or
polarization of final proton produced in collisions of unpolarized
particles will be zero identically for any kinematical conditions
of the considered reaction. Analogously, it is possible to predict
that $\rho_{11} = 1$, and all other elements of the $\rho$-meson
density matrix must be zero. Evidently contributions other than
above mechanisms should be estimated to assess the relevance of
the proposed measurement.

In consideration of other mechanisms one problem that must be
stressed is the applicability of Pomeron exchange in the near
threshold region. In accordance with resonance-Reggeon duality
\cite{R21}, at low energies sum of the resonance contributions in
s-channel can be effectively described by the different t-channel
Reggeon (but not by the Pomeron). Thus we face with double
counting problem when s-channel resonance contributions and
t-channel exchanges are considered simultaneously \cite{R22}, and
therefore division of threshold amplitude into resonance and
background cannot be done in a unique way. In this respect, Born
contributions to $\gamma+N\rightarrow\ V+N^{*}(1440)$ must be
considered as background.

Complex spin structure in matrix elements of the reaction
$\gamma+N\rightarrow\omega+N^{*}(1440)$ as compared with the
pseudoscalar meson photoproduction on nucleon have been a barrier
to go further to include the resonances. For example for spin
$J\geq 3/2$ case, there are six independent multipole amplitudes
with six unknown coupling constants different from zero and large
number of nucleon resonances $N^{*}$ in the considered reaction.
Therefore, the effects of each resonance cannot be considered with
good accuracy. Determination of V$N^{*}N^{*}$ is also another
problem in consideration of resonance mechanisms of these
reactions because there is no information directly available from
experiments. Without having the polarization data with polarized
beam, polarized target and of measurements on polarization
properties of final vector meson for the
$\gamma+N\rightarrow\omega+N^{*}(1440)$ reaction, inclusion of
resonance mechanisms do not seem as suitable for the analysis of
these reactions.

In the present work, we investigate the role played by
(s+u)-channel (N+$N^{*}(1440)$)-exchanges in,
$\gamma+N\rightarrow\ \omega +N^{*}(1440)$, associative
photoproduction of $\omega$-meson and Roper resonance in the near
threshold region ($E_{\gamma} < 3$ GeV). Our model contains
(N+$N^{*}(1440)$)-exchange mechanisms together with the
$\pi$-exchange but without Pomeron exchange. Advantage of this
model compared to the over simplified $\pi$-exchange model is that
it allows to find nonzero values for the polarization observables,
which are in T-even character, such as beam asymmetry $\Sigma$
induced by linear photon polarization , and density matrix
elements of the vector meson produced in polarized and unpolarized
particles. In the proposed model it is also possible to
discriminate the isotopic spin effects in observables on proton
and neutron targets due to $\pi\bigotimes N$-interference and
different N-contributions.

The paper is organized as follows: In Sec. II we give the model
independent formalism for calculation of differential
cross-section and beam asymmetry and describe our model in the
framework of exchange mechanisms. Results of our calculations for
the differential cross-section and beam asymmetry are presented
and are discussed in Sec. III. In the last section conclusions
extracted from the discussion of our results are given with a few
remarks.

\section{Formalism and Model}

Calculations of different observables for the associative
photoproduction $\gamma+N\rightarrow N^{*}+\omega$ are performed
by using the formalism of so called transversal amplitudes in the
center of mass system (CMS) of the considered reaction. Advantage
of this formalism is that it is effective for the analysis of
polarization phenomena in photoproduction reactions.

Matrix element of any photoproduction mechanism can be written in
terms of 12 independent transversal amplitudes as
\begin{eqnarray}
 {\cal M}& = & \varphi_{2}^{\dag}{\cal F}\varphi_{1}~~, \nonumber \\
 {\cal F}& = &
if_{1}(\vec{\varepsilon}.\hat{\vec{m}})(\vec{U}.\hat{\vec{m}})
 + if_{2}(\vec{\varepsilon}.\hat{\vec{m}})(\vec{U}.\hat{\vec{k}})
 + if_{3}(\vec{\varepsilon}.\hat{\vec{n}})(\vec{U}.\hat{\vec{n}})\nonumber
\\
 &+&(\vec{\sigma}.\hat{\vec{n}})[f_{4}(\vec{\varepsilon}.\hat{\vec{m}})
 (\vec{U}.\hat{\vec{m}})
 + f_{5}(\vec{\varepsilon}.\hat{\vec{m}})(\vec{U}.\hat{\vec{k}}) +
 f_{6}(\vec{\varepsilon}.\hat{\vec{n}})(\vec{U}.\hat{\vec{n}})]\nonumber
\\
 &+&(\vec{\sigma}.\hat{\vec{m}})[f_{7}(\vec{\varepsilon}.\hat{\vec{m}})
 (\vec{U}.\hat{\vec{n}})
 + f_{8}(\vec{\varepsilon}.\hat{\vec{n}})(\vec{U}.\hat{\vec{m}}) +
 f_{9}(\vec{\varepsilon}.\hat{\vec{n}})(\vec{U}.\hat{\vec{k}})]\nonumber
\\
 &+&(\vec{\sigma}.\hat{\vec{k}})[f_{10}(\vec{\varepsilon}.\hat{\vec{m}})
 (\vec{U}.\hat{\vec{n}})
 + f_{11}(\vec{\varepsilon}.\hat{\vec{n}})(\vec{U}.\hat{\vec{m}})
 + f_{12}(\vec{\varepsilon}.\hat{\vec{n}})(\vec{U}.\hat{\vec{k}})]~~,
\end{eqnarray}
where $\hat{\vec{m}}$, $\hat{\vec{n}}$, and $\hat{\vec{k}}$ are
defined as $\hat{\vec{k}}=\vec{k}/|\vec{k}|$, $
\hat{\vec{n}}=\vec{k}\times\vec{q}/ |\vec{k}\times\vec{q}|$, $
\hat{\vec{m}}= \hat{\vec{n}}\times\hat{\vec{k}}$, $\vec{k}$ and
$\vec{q}$ are the three-momentum of the photon and the vector
meson in CMS, $\varphi_{1}$($\varphi_{2}$) is the two-component
spinor for initial nucleon and final Roper resonance, and
transversal amplitudes $f_{i}$, $i=1,...,12$, are complex
functions of s and t, $f_{i}=f_{i}(s,t)$.

Differential cross section and beam asymmetry are given by
\begin{eqnarray}
 \frac{d\sigma}{d\Omega}=\frac{1}{2}{\cal N}\overline{{\cal F}{\cal
F}^{\dagger}}
\end{eqnarray}
and
\begin{equation}
\Sigma=\frac{d\sigma_{\parallel}/d\Omega-d\sigma_{\perp}/d\Omega}{d\sigma_
 {\parallel}/d\Omega+d\sigma_{\perp}/d\Omega}~~,
\end{equation}
where ${\cal N}=|\vec{q}|/64\pi^{2} s |\vec{k}|$ and
$d\sigma_{\parallel}/d\Omega$ ($d\sigma_{\perp}/d\Omega$) is the
differential cross section induced by photon whose polarization is
parallel (perpendicular) to reaction plane in which all other
particles in the initial and final state are unpolarized. The
corresponding differential cross section and beam asymmetry which
are obtained by using Eq. (1), (2), and (3) can be written in
terms of transversal amplitudes $f_{i}$
\begin{eqnarray}
 \frac{d\sigma}{d\Omega}& = &{\cal N}\left(h_{1}+h_{2}\right)~~,\nonumber
\\
 \Sigma& = & \frac{\left(h_{1}-h_{2}\right)}{\left(h_{1}+h_{2}\right)}~~,
 \nonumber \\
 h_{1}&=&
 \frac{1}{2}\{\left[|f_{1}|^{2}+|f_{2}|^{2}+|f_{4}|^{2}+|f_{5}|^{2}
 +|f_{7}|^{2} +|f_{10}|^{2}\right]\nonumber \\
 &+&\left[\frac{|\vec{q}|^{2}\sin^{2}\theta}{m_{v}^{2}}\right]\left[|f_{1}|^{2}
 +|f_{4}|^{2}\right]+\left[\frac{|\vec{q}|^{2}\cos^{2}\theta}{m_{v}^{2}}\right]
 \left[|f_{2}|^{2}+|f_{5}|^{2}\right]\nonumber\\
 &+&\left[\frac{|\vec{q}|^{2}2\sin\theta\cos\theta}{m_{v}^{2}}\right]Re\left[(f_{1}
 f_{2}^{*})+(f_{4}f_{5}^{*})
 \right]\}~~,\nonumber\\
 h_{2} &=&
 \frac{1}{2}\{\left[|f_{2}|^{2}+|f_{6}|^{2}+|f_{8}|^{2}+|f_{9}|^{2}
 +|f_{11}|^{2}+|f_{12}|^{2}\right]\nonumber\\
 &+&\left[\frac{|\vec{q}|^{2}\sin^{2}\theta}{m_{v}^{2}}\right]\left[|f_{8}|^{2}
 +|f_{11}|^{2}\right]+
 \left[\frac{|\vec{q}|^{2}\cos^{2}\theta}{m_{v}^{2}}\right]\left[|f_{9}|^{2}+|f_{12}|
 ^{2}\right]\nonumber \\
 &+&\left[\frac{|\vec{q}|^{2}2\sin\theta\cos\theta}{m_{v}^{2}}\right]Re\left[(f_{8}
 f_{9}^{*})+(f_{11}f_{12}^{*})
 \right]\}~~,
\end{eqnarray}
where $m_{v}$ is the mass of vector meson, $\theta$ is the angle
between $\vec{k}$ and $\vec{q}$ in CMS, $h_{1}$ and $h_{2}$ are
the structure functions of the considered reaction.

Due to large width of Roper resonance, width effects must be
included in the calculation of $\gamma p\rightarrow\omega
N^{*}(1440)$ reaction near threshold. We introduce these effects
by the Breit-Wigner parametrization as
\begin{equation}
 \frac{d\sigma}{d\Omega}(\gamma p\rightarrow\omega
 N^{*}(1440)\rightarrow\omega\pi^{0} p)=\int_{m_{\pi^{0}}+m_{p}}
 ^{M_{N^{*}}^{max}(s,t)}\frac{d\sigma}{dt}(\gamma p\rightarrow\omega
 N^{*}(1440))B(M_{N^{*}})dM_{N^{*}}
\end{equation}
in which $M_{N^{*}}^{max}$ is the maximum mass of the Roper
resonance for fixed s and t, and $B(M_{N^{*}})$ is the
Breit-Wigner function in the form
\begin{eqnarray}
 B(M_{N^{*}})=\frac{2}{\pi}\frac{M_{N^{*}}M_{N^{*}}^{0}
 \Gamma_{N^{*}(1440)\rightarrow\pi^{0} p}(M_{N^{*}})}{(M{_{N^{*}}^{0~2}}
 -M_{N^{*}}^{2})^{2}
 +M{_{N^{*}}^{0~2}}~\Gamma_{N^{*}(1440)}^{2}(M_{N^{*}})}~.
\end{eqnarray}
Energy dependent partial and total widths are given by
\begin{eqnarray}
 \Gamma_{N^{*}(1440)\rightarrow\pi^{0} p (M_{N^{*}})}&=&
 \Gamma_{N^{*}(1440)\rightarrow\pi^{0} p} (M_{N^{*}}^{0})\nonumber \\
 &&\frac{M_{N^{*}}^{0}}{M_{N^{*}}}\left[\frac{E(M_{N^{*}}-M_{N})}
 {E(M_{N^{*}}^{0}-M_{N})}\right] \frac{p[E(M_{N^{*}}]}{p[E(M_{N^{*}}^{0}]}
\end{eqnarray}
and
\begin{eqnarray}
 \Gamma_{N^{*}(1440)}^{tot}(M_{N^{*}})&=&
 \Gamma_{N^{*}(1440)}(M_{N^{*}}^{0})\nonumber \\
 &&\frac{M_{N^{*}}^{0}}{M_{N^{*}}}\left[\frac{E(M_{N^{*}}-M_{N})}
 {E(M_{N^{*}}^{0}-M_{N})}\right]
 \frac{p[E(M_{N^{*}}]}{p[E(M_{N^{*}}^{0}]}~,
\end{eqnarray}
where $E(M_{N^{*}}(p(E(M_{N^{*}}))$ is the energy(three momentum)
of $M_{N^{*}}$ in the rest frame of decay
$N^{*}(1440)\rightarrow\pi N$. We use the values for
$\Gamma_{N^{*}(1440)\rightarrow\pi^{0} p} (M_{N^{*}})=76$~MeV and
$\Gamma_{N^{*}(1440)}^{tot}(M_{N^{*}}^{0})=350$~MeV \cite{R1}.

The suggested model for the reaction $\gamma+N\rightarrow
N^{*}(1440)+\omega$ contains t-, s-, and u-channel exchange
mechanisms, which are shown in Fig. 1. Following discussion of Ref
\cite{R19} we consider only $\pi$-exchange mechanism in t-channel.
The matrix element for this exchange mechanism can be written as
\begin{equation}
 {\cal M}_{t}
 =e~\frac{g_{\omega\pi\gamma}}{m_{\omega}}~\frac{g_{\pi
NN^{*}}}{t-m_{\pi}^{2}}
 F_{\pi
NN^{*}}(t)~F_{\omega\pi\gamma}(t)~(\overline{u}(p_{2})~\gamma_{5}~u(p_{1}))
 ~(\epsilon^{\mu\nu\alpha\beta}
 ~\varepsilon_{\mu}~k_{\nu}~U_{\alpha}~q_{\beta})~,\nonumber \\ \nonumber
\\
\end{equation}
where $t=(k-q)^{2}$, $m_{\omega}(m_{\pi})$ is the mass of the
$\omega$-($\pi$-) meson, $\varepsilon_{\mu}(U_{\mu})$ is the
polarization four vector of photon(vector meson),
$u(p_{1})(u(p_{2}))$ is Dirac spinor for initial nucleon (final
Roper resonance), $g_{\pi NN^{*}}$ and $g_{\omega\pi\gamma}$ are
the strong and electromagnetic coupling constants of the $\pi
NN^{*}$ and $\omega\pi\gamma$ vertices, respectively. Notation of
particle four momenta is given in Fig. 1. Form factors which
appear in the above matrix element are in the form
\begin{eqnarray}
F_{\pi NN^{*}}(t)=\frac{\Lambda_{\pi NN^{*}}^2-m_{\pi}^2}{\Lambda_{\pi
NN^{*}}^2-t}~~,
F_{\omega\pi\gamma}(t)=\frac{\Lambda_{\omega\pi\gamma}^2-m_{\pi}^2}
{\Lambda_{\omega\pi\gamma}^2-t},
\end{eqnarray}
where $\Lambda_{\pi NN^{*}}$($\Lambda_{\omega\pi\gamma}$) is the
cut-off parameter of the considered vertices in pole diagrams.

The nucleon s-channel contribution is described by the following
amplitude
\begin{eqnarray}
{\cal M}_{s}& = & \frac{e}{s-M^{2}}
\overline{u}(p_{2})(g_{VNN^{*}}^{V}\hat{U}+
\frac{g_{VNN^{*}}^{T}}{(M+M^{*})}
\hat{U}~\hat{q})(\hat{p_{1}}+\hat{k}+M) (Q_{N} \hat{\varepsilon}-
\frac{\kappa_{N}}{2 M} \hat{\varepsilon}~\hat{k})u(p_{1})\nonumber \\
\end{eqnarray}
where $\varepsilon\cdot k=$ $U\cdot q=0$,
$\hat{a}=\gamma^{\mu}a_{\mu}$, $M(M^{*})$ is the mass of initial
nucleon (final Roper resonance), $Q_{N}=1(0)$ is the electric
charge for proton (neutron), $\kappa_{N}=1.79(-1.91)$ is the
anomalous magnetic moment for proton (neutron). Different from the
nucleon exchange in s-channel amplitude, Roper resonance exchange
in u-channel amplitude is considered which is given by
\begin{eqnarray}
{\cal M}_{u}= \frac{e}{u-M{^{*}}^{2}} \overline{u}(p_{2})(Q_{N}
\hat{\varepsilon}- \frac{\kappa_{N}^{*}}{2 M^{*}}
\hat{\varepsilon}~\hat{k})(\hat{p_{2}}-\hat{k}+M^{*}) (g_{\omega
NN^{*}}^{V}\hat{U}+
\frac{g_{\omega NN^{*}}^{T}}{(M+M^{*})} \hat{U}~\hat{q}) u(p_{1})
\end{eqnarray}
where $u=(k-p_{2})^{2}$, $\kappa_{N}^{*}$ is the anomalous
magnetic moment of Roper resonance $N^{*}(1440)$. Neglecting their
possible dependence on the virtuality in s and u of the
intermediate nucleon and Roper resonance,  $g_{\omega NN^{*}}^{V}$
and $g_{\omega NN^{*}}^{T}$ (vector and tensor coupling constants)
are chosen to be the same in both channel matrix elements.

For s- and u-channel amplitudes it is possible to dress the form
factors with s and u dependencies either in the form of F(s) and
F(u) or  F(s,u). Use of the form factor as in the first case
causes the violation of the gauge invariance. Even if the latter
form preserve the gauge invariance this type of phenomenological
form factor \cite{R23}, being the function of both Mandelstam
variables, behaves like an amplitude rather than form factor.
Therefore, following the prescription of the Ref. \cite{R24}, we
use the constant form factors $F(s)=F(u)=1$. In this case the
effects are absorbed by the coupling constants $g_{\omega
NN^{*}}^{V}$ and $g_{\omega NN^{*}}^{T}$ of $\omega
NN^{*}$-vertices.

Let us note that these couplings are free parameters of our model
and their values must be different from the values in space-like
region of the vector meson momentum. In literature the values of
the $g_{\omega NN^{*}}^{V}$ and $g_{\omega NN^{*}}^{T}$ coupling
constants are obtained from the reactions $N+N\rightarrow N^{*}+N$
and $N^{*}+N\rightarrow N+N$ \cite{R25} following arguments of
Ref.\cite{R26}. Hovewer, the values of such coupling constants
obtained from the NN- and $NN^{*}$- potentials will be different
from that of coupling constants used in the associative
photoproduction of $\omega$-meson and Roper resonance because they
are considered in different regimes, space-like in the first case
whereas time-like in the latter case. Another approach in
calculation of transition couplings for virtual meson is suggested
in the framework of constituent quark model \cite{R27}, but the
values obtained are suggestive rather than being definite
quantitative predictions. At this stage determining the values of
$\omega NN^{*}(1440)$ coupling constants is also not possible due
to the fact that there is no direct experimental data on these
coupling strengths. To overcome this problem we assume that the
values of $\omega NN^{*}(1440)$ coupling constants are equal to
that of $\omega NN$ coupling constants. With this assumption it is
possible to determine the values of these constants from the fit
to experimental data on the differential cross-section for the
photoproduction of $\omega$ meson \cite{R28}.

\section{Results and Discussion}

In the previous section, we have defined all the necessary
parameters in t-, s-, and u-channel amplitudes of our model for
the process, $\gamma+N\rightarrow\omega+N^{*}(1440)$. Let us
specify here in more detail the coupling strenghts and cut-off
parameters of of the considered model. For the  coupling constant
$g_{\omega \pi \gamma}$ we take the most commonly used value
$1.82$ \cite{R19}obtained from the experimental partial decay
width of $\omega \rightarrow \pi \gamma$ decay. The situation is
however not clear for coupling strength of $\pi
NN^{*}(1440)$-vertex. Because of the large uncertainity in the
partial decay width of the $N^{*}(1440)$ into $N \pi$ channel
$(228\pm 82 MeV)$ the coupling constant $g_{\pi NN^{*}(1440)}$
cannot be determined precisely. Following \cite{R20} we will use
the value $3.4$ for the $g_{\pi NN^{*}(1440)}$.

The remaining inputs of our model are the $\omega NN^{*}(1440)$
coupling constants and cut-off parameters $\Lambda_{i}$. In
consideration of the coupling constants not only their absolute
values but also their relative signs are important because of the
essential interference effects. Cut-off parameters are in any case
positive and by convention $g_{\omega \pi \gamma}g_{\pi
NN^{*}(1440)}$ is chosen as positive. Therefore, the signs of
$g_{\omega NN^{*}(1440)}^{V}$ and $g_{\omega NN^{*}(1440)}^{ T}$
that appear in our results are their relative signs with respect
to the $\pi$-contribution, and not their absolute signs. Since
applicability of the same form factors for different precesses is
not proved rigorously we can fixed  absolute values of cut-off
parameters at some plausible values. Consequently, we are left
with two fitting parameters $g_{\omega NN^{*}}^{V}$ and $g_{\omega
NN^{*}}^{T}$.

For our calculations, the following three different sets with
almost the same value of $\chi^{2}$, which are obtained from the
fit to the experimental data about $d\sigma (\gamma p\rightarrow
p\omega )/dt$ in the near threshold region \cite{R28}, are chosen
for the coupling constants $g_{\omega NN^{*}}^{V,T}$:

\underline{Set 1:}
\begin{equation}
 g_{\omega NN^{*}}^{V}=-1.4,~~ g_{\omega NN^{*}}^{T}=0.4,~~\chi^{2}=2.2
 \nonumber \\
\end{equation}

\underline{Set 2:}
\begin{equation}
 g_{\omega NN^{*}}^{V}=0.5,~~ g_{\omega NN^{*}}^{T}=0.1,~~\chi^{2}=1.6
 \nonumber \\
\end{equation}

\underline{Set 3:}
\begin{equation}
 g_{\omega NN^{*}}^{V}=-0.01,~~ g_{\omega NN^{*}}^{T}=0.6,~~\chi^{2}=1.9
 \nonumber \\
\end{equation}

To obtain set 1, we use the standard values of cut-off parameters
$\Lambda_{\pi NN^{*}}=\Lambda_{\pi NN}=0.7$ GeV and
$\Lambda_{\omega\pi\gamma}=0.77$ GeV. In analyzing the sensitivity
of the best fit to $\Lambda_{\pi NN^{*}}$ and
$\Lambda_{\omega\pi\gamma}$, we discover that the standard values
of $\Lambda_{i}$ do not give the best solution. If the values of
$\Lambda_{\pi NN^{*}}$ and $\Lambda_{\omega\pi\gamma}$ are changed
to 0.5 GeV and 1.0 GeV, respectively, we find better sets for the
coupling constants $g_{\omega NN^{*}}^{V, T}$, namely set 2 and
set 3. We follow the same minimization procedure used in Ref.
\cite{R29} for the determination of vector and tensor coupling
constant values.

Differential cross section for $\gamma p\rightarrow\omega
N^{*}(1440)$ reaction at $E_{\gamma}=2.5$ GeV using the above sets
of coupling constants of our model and zero-width approximation is
shown in the left panel of Fig. 2. All these sets give different
cross section for $\gamma p\rightarrow\omega N^{*}(1440)$. We also
consider the width effects of Roper resonance on differential
cross section, assuming that $N^{*}(1440)$ decays subsequently
into the $\pi^{0} p$ channel. These effects are presented in the
right panel of Fig. 2. At $-t=0.36$ $GeV^{2}$, differential cross
section for the process $\gamma p\rightarrow\omega
N^{*}(1440)\rightarrow\omega\pi^{0} p$ is about 20 times smaller
than differential cross section for $\gamma p\rightarrow\omega
N^{*}(1440)$ in the zero-width approximation. This difference
comes from the partial decay width of $N^{*}(1440)$ into $\pi^{0}
p$ channel which is nearly 20{\%%
} and interval of the $M_{N^{*}}$ appear in the integral of Eq.
(5) reduces the strength of Roper resonance excitation by a factor
of about 4 compared to case where all strength is concentrated at
$M_{N^{*}}^{0}=1.44$ GeV. Progress on the width effects directly
is linked to the availability of new experimental data providing
constraints on the couplings of $N^{*}(1440)$ to the $\pi N$ and
$\omega N$ channels.

The contributions of different amplitudes to $d\sigma(\gamma
p\rightarrow\omega N^{*}(1440))/dt$ and $d\sigma(\gamma
n\rightarrow\omega N^{*}(1440))/dt$ are presented in Fig. 3 and 4.
Set 1 and set 3 for the coupling constants produce negative
$\pi\bigotimes(N+N^{*}(1440))$-interference while set 2 has a
positive interference in the differential cross section for the
associative photoproduction of Roper resonance and $\omega$-meson
on proton and neutron targets. For all these cases up to $-t=0.5$
our predictions for differential cross section does not differ
significantly from the one-pion exchange results, but beyond this
value of t predictions of both model are different. Predicted
behaviour of differential cross section for $\gamma
n\rightarrow\omega N^{*}(1440)$ as compared with $\gamma
p\rightarrow\omega N^{*}(1440)$ indicates that differential cross
section on proton and neutron targets can have differences by a
factor of 2 or more, i.e. we can predict definite isotopic
effects.

Another prediction of our model is the t-dependence of beam
asymmetry $\Sigma(\gamma p\rightarrow\omega N^{*}(1440))$ and
$\Sigma(\gamma n\rightarrow\omega N^{*}(1440))$ at
$E_{\gamma}=2.5$ GeV, shown in Fig. 5 and Fig. 6. For the proton
target, all three sets of coupling constants produce negative
$\Sigma$, but although absolute value of $\Sigma$ is small for set
1 and set 2, being $|\Sigma|\leq 0.1$, it is nearly 0.25 at
$|t|=1.2$ $GeV^{2}$. However, in the neutron case, especially for
set 1, beam asymmetry shows different behaviour, i.e positive in
sign for $|t|\leq 0.6$ $GeV^{2}$ and negative in the rest of the
interval of $|t|$. Moreover, our model results show that beam
asymmetry is sensitive to the sets of coupling constants in our
model.

At present time, there is no systematic investigation of the role
played by the (s+u)-contribution in associative photoproduction of
the Roper resonance and $\omega$ meson in near threshold region.
In fact, the unknown $\omega NN^{*}(1440)$ couplings have been the
barrier to go further to include this contributions. However, our
approach to determine these couplings make detail analysis
possible for the description differential cross section and beam
asymmetry. At this stage, of course, it is very difficult to say
that our results are decisive because of the absence of the any
differential cross section and polarization  data about the
processes $\gamma N\rightarrow N^{*}(1440) \omega$, but we can
test our model by comparing it with the proposed one-boson
exchange model, which  include only $\pi$-contribution and is
valid in the region $|t|\leq 0.5-0.6~GeV^{2}$. In this region our
predictions of the differential cross section for $\gamma
p\rightarrow \omega N^{*}(1440)$ are consistent with the
predictions of Ref. \cite{R20} obtained from $\pi$-exchange model.
This indicates that if the simple $\pi$-exchange model make sense
our assumption about coupling strengths of $\omega NN^{*}(1440)$
is reasonable. Therefore, this model seems appropriate to perform
the calculations on the boundary of the modern approaches to these
processes and it should be considered as a first approach.

\section{Conclusions}

The analysis done in the previous section results in the following
conclusions:

$\bullet$ The relatively simple model $(\pi +N+N^{*})$ is proposed
to describe the assocative photoproduction of Roper resonance and
$\omega$ meson on proton and neutron targets near threshold region
$(E_{\gamma}<3 GeV)$ for the whole t region. Comparison of this
model with one-pion calculations demonstrates the definite
difference in behaviour of differential cross section.

$\bullet$ The different solutions for the coupling constants and
the cut-off parameters obtained from the fitting procedure result
in the constructive and destructive $\pi \bigotimes
(N+N^{*})$-interference contributions to $d\sigma (\gamma
p\rightarrow N^{*}(1440))/dt$.

$\bullet$ $\Sigma$-asymmetry is different from zero and its
t-behaviour is sensitive to  our model parameters, namely
$g_{\omega NN^{*}(1440)}^{V,T}$ coupling constants, which are
obtained in the time-like region of vector meson four momentum.

\section*{Acknowledgments}

We thank M. P. Rekalo for suggesting this problem to us and
gratefully acknowledge his guidance during the course of our work.
We also thank to our supervisors A. G\7kalp and O. Y\5lmaz for
their contributions and continuous attentions.

\newpage
\begin{figure}
$\left. \right.$
\vskip 15cm
    \includegraphics{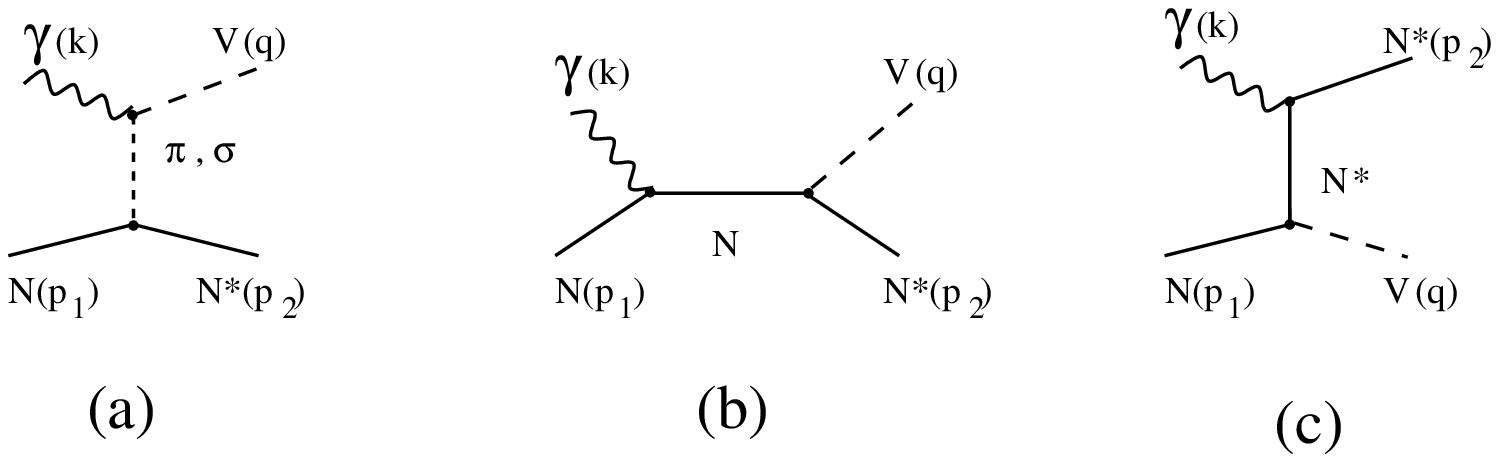} \vskip -1cm \caption{Mechanisms of
the model for assocative photoproduction of Roper resonance and
$\omega$-photoproduction: (a) t-channel exchanges, (b) and (c) s-
and u-channel nucleon exchanges. }
\end{figure}
\newpage
\begin{figure}
$\left. \right.$
\vskip 10cm
    \includegraphics{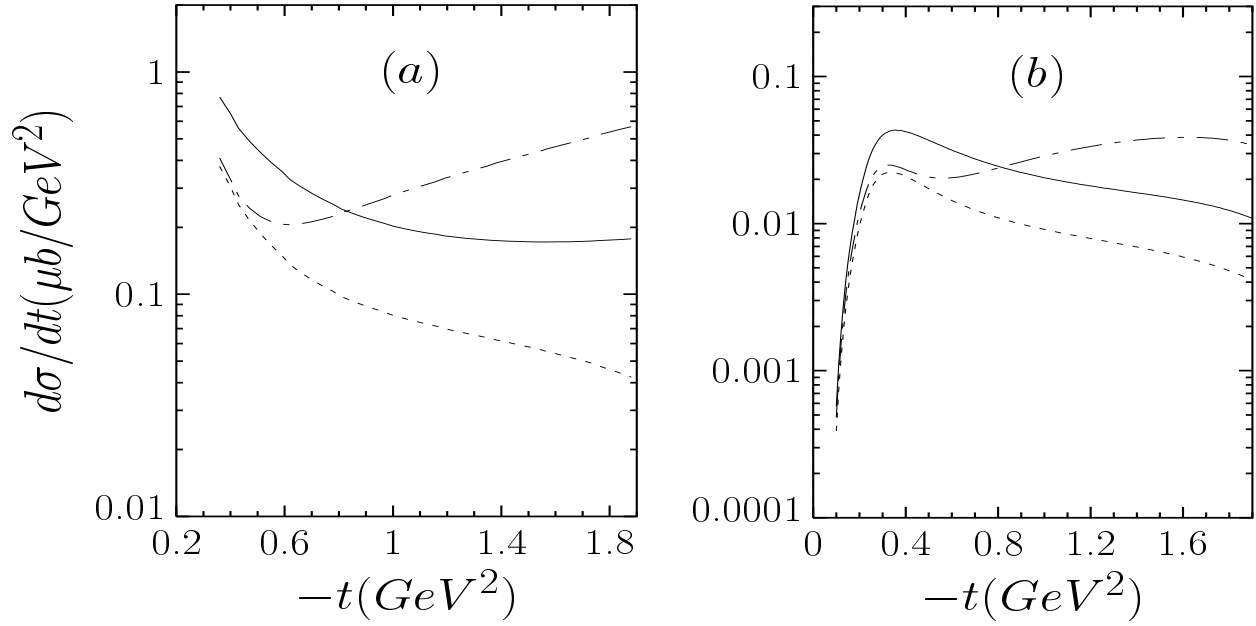}
\vskip 5.5cm
\caption{Differential cross section for the processes, (a) $\gamma
p\rightarrow\omega N^{*}(1440)$ in the zero-width approximation
(b) $\gamma p\rightarrow\omega N^{*}(1440)\rightarrow\omega\pi^{0} p$,
at $E_{\gamma}=$ $2.5$~GeV. Solid, dashed and dot-dashed lines correspond
to $g_{\omega NN}^{V}=0.5$ and $g_{\omega NN}^{T}=0.1$; $g_{\omega
NN}^{V}=-0.01$ and
$g_{\omega NN}^{T}=0.6$; $g_{\omega NN}^{V}=-1.4$,
$g_{\omega NN}^{T}=0.4$, respectively.}
\end{figure}
\newpage
\begin{figure}
$\left. \right.$
\vskip 10cm
    \includegraphics{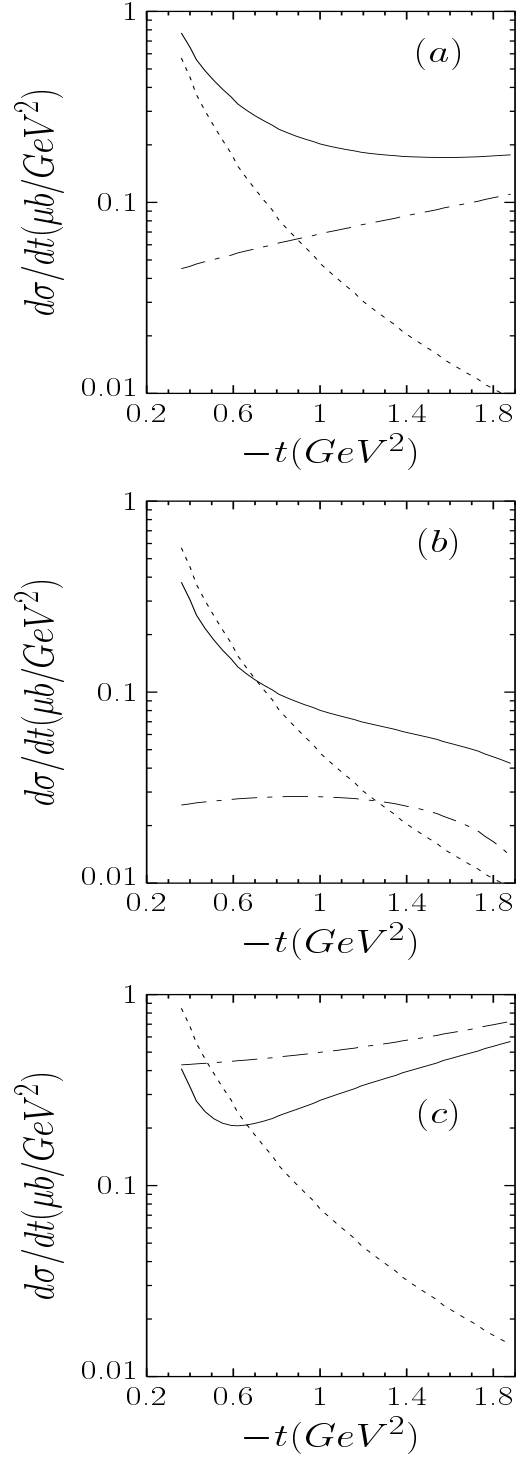} \vskip 9.5 cm \caption{Different
contributions to $d\sigma(\gamma p\rightarrow \omega
N^{*}(1440))/dt$ at $E_{\gamma}=$ $2.5$~GeV for three different
fitted parameter values: (a)  $g_{\omega NN}^{V} =0.5$,
~$g_{\omega NN}^{T}=0.1$, (b) $g_{\omega NN}^{V}=-0.01$,
~$g_{\omega NN}^{T}=0.6$, and (c) $g_{\omega NN}^{V}=-1.4$,
~$g_{\omega NN}^{T}=0.4$. Solid, dashed and dot-dashed lines
correspond to total, $\pi$-, and (s+u)-contribution,
respectively.}
\end{figure}
\newpage
\begin{figure}
$\left. \right.$
\vskip 10cm
    \includegraphics{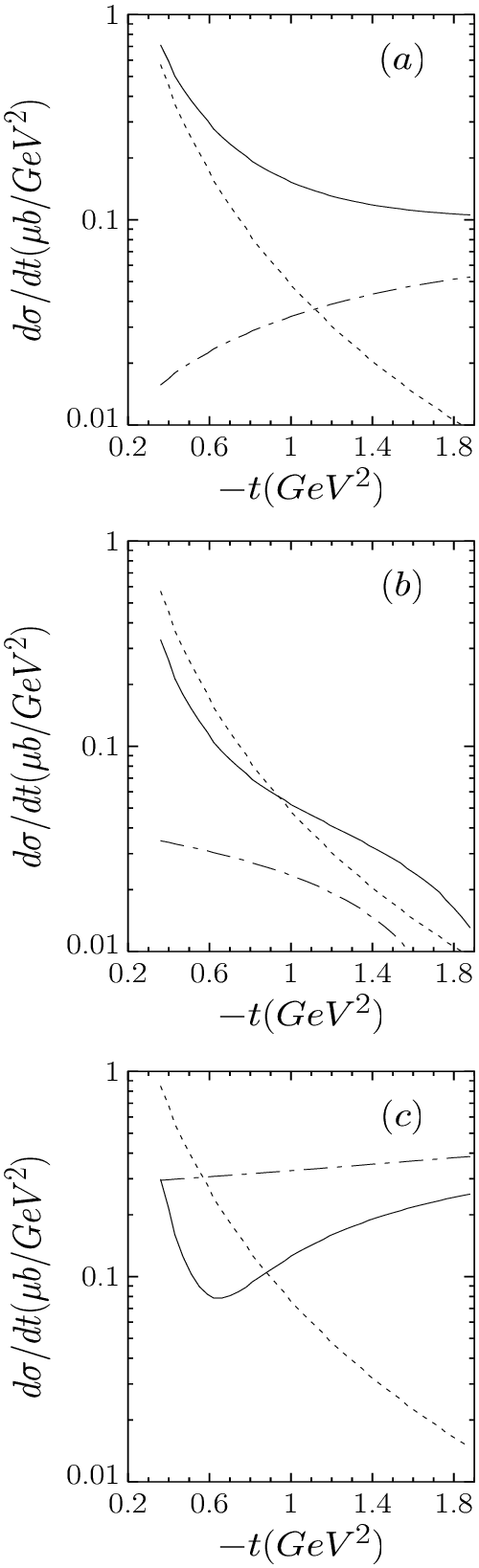} \vskip 10.0 cm \caption{Different
contributions to $d\sigma(\gamma n\rightarrow \omega
N^{*}(1440))/dt$ at $E_{\gamma}=$ $2.5$~GeV for three different
fitted parameter values: (a)  $g_{\omega NN}^{V} =0.5$,
~$g_{\omega NN}^{T}=0.1$, (b) $g_{\omega NN}^{V}=-0.01$,
~$g_{\omega NN}^{T}=0.6$, and (c) $g_{\omega NN}^{V}=-1.4$,
~$g_{\omega NN}^{T}=0.4$. Notations are same as in Fig. 3.}
\end{figure}
\newpage
\begin{figure}
$\left. \right.$
\vskip 10cm
    \includegraphics{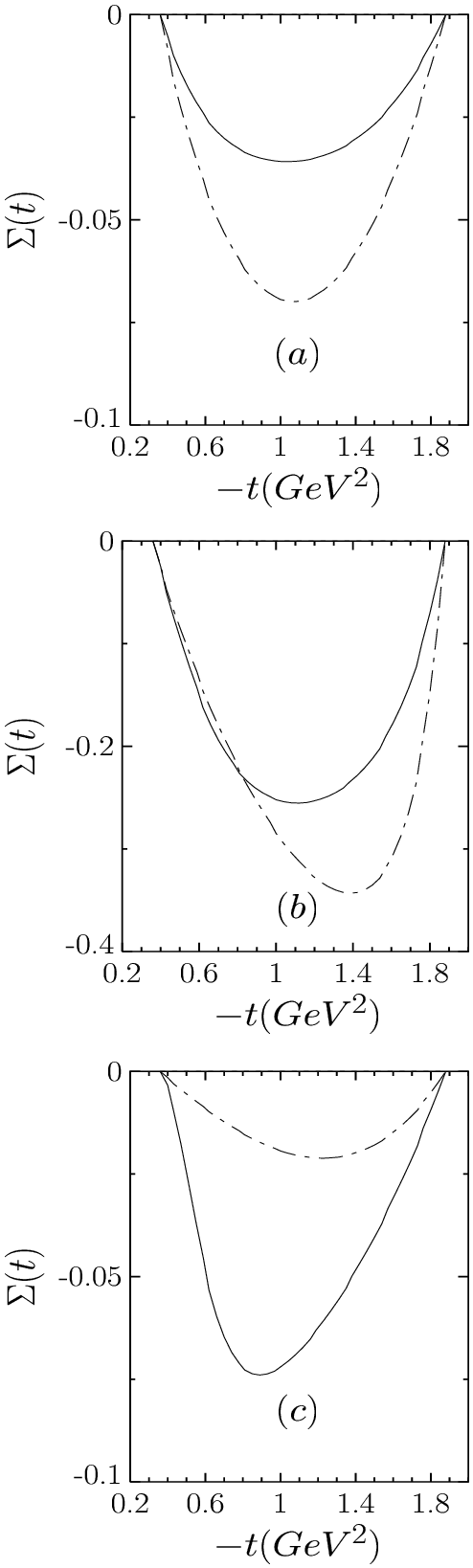} \vskip 10.0 cm \caption{Different
contributions to $\Sigma(\gamma p\rightarrow \omega N^{*}(1440))$
at $E_{\gamma}=$ $2.5$~GeV for three different fitted parameter
values: (a)  $g_{\omega NN}^{V} =0.5$, ~$g_{\omega NN}^{T}=0.1$,
(b) $g_{\omega NN}^{V}=-0.01$, ~$g_{\omega NN}^{T}=0.6$, and (c)
$g_{\omega NN}^{V}=-1.4$, ~$g_{\omega NN}^{T}=0.4$. Notations are
same as in Fig. 3.}
\end{figure}
\newpage
\begin{figure}
$\left. \right.$
\vskip 10cm
    \includegraphics{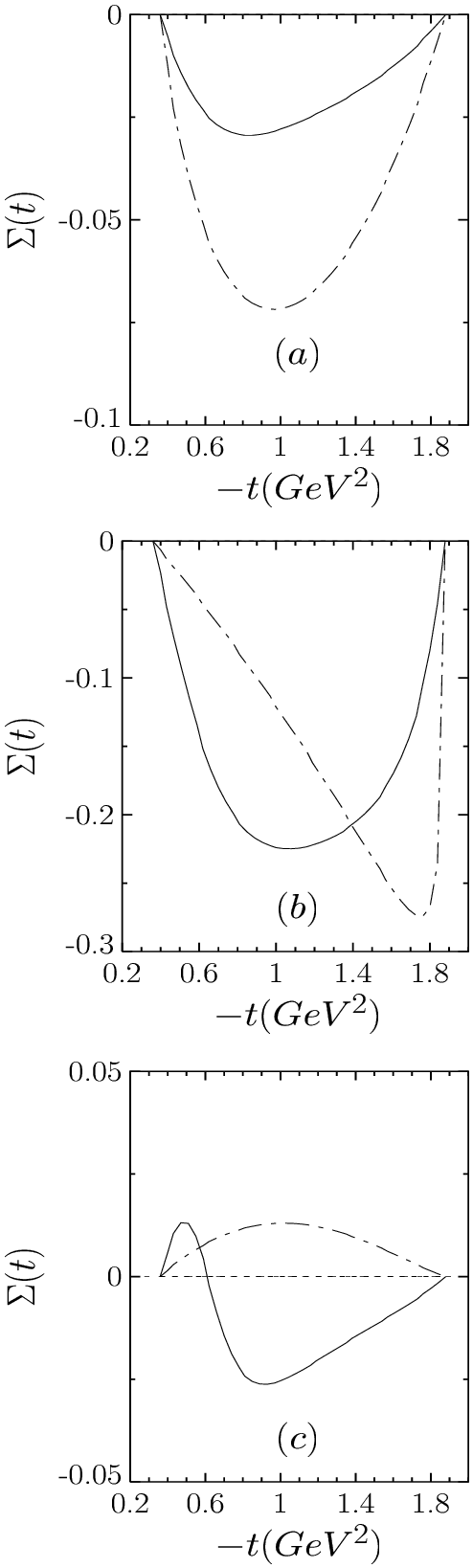} \vskip 10.0 cm \caption{Different
contributions to $\Sigma(\gamma n\rightarrow \omega N^{*}(1440))$
at $E_{\gamma}=$ $2.5$~GeV for three different fitted parameter
values: (a)  $g_{\omega NN}^{V} =0.5$, ~$g_{\omega NN}^{T}=0.1$,
(b) $g_{\omega NN}^{V}=-0.01$, ~$g_{\omega NN}^{T}=0.6$, and (c)
$g_{\omega NN}^{V}=-1.4$, ~$g_{\omega NN}^{T}=0.4$. Notations are
same as in Fig. 3.}
\end{figure}
\end{document}